\documentclass[useAMS,usenatbib]{pasa}

\title[Beyond the galaxy luminosity function]{Beyond the galaxy luminosity function}
\author[Driver]{Simon Driver}
\affil{Mount Stromlo Observatory, Weston, ACT 2902, AUSTRALIA
spd@mso.anu.edu.au}

\begin{document}

\maketitle

\label{firstpage}

\begin{abstract}
With the advent of large scale surveys ({\it i.e.,} Legacy Surveys) it
is now possible to start looking beyond the galaxy luminosity function
(LF) to more detailed statistical representations of the galaxy
population, {\it i.e.,} multivariate distributions. In this review I
first summarise the current state-of-play of the $B$-band global and
cluster LFs and then briefly present two promising bivariate
distributions: the luminosity-surface brightness plane (LSP); and the
colour-luminosity plane (CLP). In both planes galaxy bulges and galaxy
disks form marginally overlapping but distinct distributions,
indicating two key formation/evolutionary processes (presumably merger
and accretion). Forward progress in this subject now requires the
routine application of reliable bulge-disk decomposition codes to
allow independent investigation of these two key components.
\end{abstract}



\section{Introduction}
\begin{figure}
\includegraphics{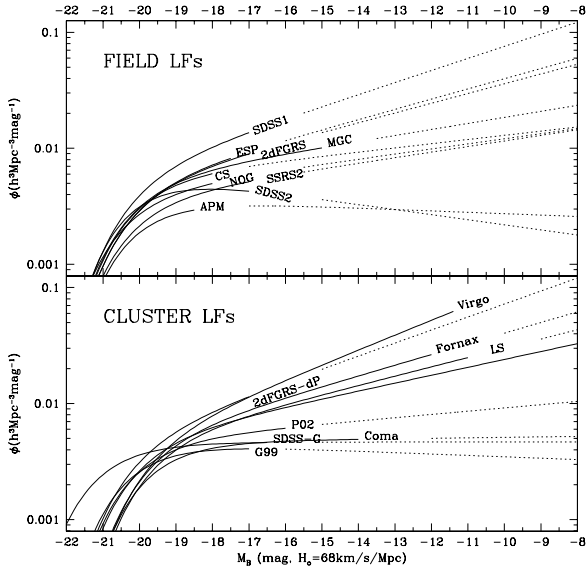}
\vspace*{75mm}
\caption{Various luminosity functions as measured for the global
environment (upper) and cluster environment (lower). Data are taken
from Table.~3 of Liske et al (2003), Table.~1 of Driver \& De Propris
(2003), Table.~2 \& 3 from Blanton et al (2003) and Table.~2 from
Driver et al (2004). The solid lines show the regions over which the
luminosity functions have been fitted and the dotted lines the
extrapolations. The cluster luminosity functions have all been
arbitrarily normalised to $\phi^*=0.0161$ galaxies per $h^3$
Mpc$^{-3}$.  The main point is that until the systematics are resolved
one cannot draw any reasonable inference other than the field and
cluster LFs are broadly consistent. \label{fig1} }
\end{figure}

For almost 30 years the Schechter luminosity function (LF;
\citealp{schechter}) has been the standard tool for quantifying the
galaxy population\footnote{The Schechter function:
$d(\phi)=\phi^*(\frac{L}{L^*})^{\alpha}
e^{(-\frac{L}{L^*})}d(\frac{L}{L^*})$ has three key parameters, $L^*$
the characteristic luminosity where the exponential cutoff cuts in,
$\phi^*$, the normalisation at this characteristic luminosity, and
$\alpha$, the faint-end slope parameter. A value of $\alpha= -1$
implies equal numbers of galaxies in magnitude intervals, a more
negative (or steep) value implies numerous dwarf systems.}. The LF is
loosely based on the Press-Schechter formalisation for the primordial
halo distribution \citep{PS74}.  Moreover the LF consistently provides
a good formal fit to the observed luminosity distribution (LD; see for
example \citealp{norberg02}). This consistency, between the LD and LF,
appears to hold regardless of environment (\citealp{rdp2003};
\citealp{driver03}). The only departure from a pure Schechter function
appears to be in the central cores of rich clusters, where the galaxy
LD is often seen to show a marked upturn at the giant-dwarf boundary
($M_B \approx -16$ mag). Perhaps the most well known example is the
central LD of the Coma cluster (e.g., \citealp{trentham};
\citealp{beijer2002}; \citealp{andreon} and references therein). The
most plausible explanation is that the core contains an overdensity of
giant and dwarf ellipticals bolstering both the bright and faint-end
of the core cluster LF. For example the more extensive Coma survey by
\cite{mobasher2003} recovers a flat and invariant LD/LF ($\alpha=-1$)
to $M_{B} \approx -14$ mag. The phenomena of an upturn in the LD, has
also been seen in Virgo (\citealp{bothun}; \citealp{trent2002}), A963
(\citealp{a963}), A868 (\citealp{a868}), A2554 (\citealp{smith}) and
A2218 (\citealp{pracey}) for example. However, apart from these
``active'' core environments, the overall LDs from the field, to the
local group, to the local sphere, and near \& far rich clusters all
consistently follow a smooth LF within the luminosity ranges
probed. Fig.~\ref{fig1} shows an (incomplete) summary of $b,B,V$ or
$g$-band field and cluster LFs colour corrected to the Johnson B
filter.

The main point to take from Fig.~\ref{fig1} is that the global and
cluster LFs each show a broad but overlapping range of distributions.
Clearly one cannot reasonably argue for any significant variation
between the global and overall cluster environment on the basis of
these data. Studies based within the same survey data, for example
the two-degree field galaxy redshift survey study by \cite{croton},
generally find fairly subtle changes with environment. Hence it seems
that the variations seen in Fig.~\ref{fig1} indicates an
unspecified systematic error in the various studies. The most lauded
of these is the unsavoury topic of surface brightness selection
effects (\citealp{disney}, \citealp{impey97}). The concern is that the
galaxy population at each luminosity interval occupies a range in
surface brightness (or size). Surveys with shallow detection isophotes
may miss both light from a galaxy's halo, as well as entire galaxies
(see for example \citealp{spray} and
\citealp{dalcanton}). \cite{cross02} explored this possibility in
detail and demonstrated that indeed surface brightness selection
effects can play havoc with the recovered Schechter function
parameters and reproduce exactly the kind of variation seen in both
the global and cluster LFs of Fig.~\ref{fig1}.

More recently a number of papers have identified a clear
luminosity-surface brightness (or size)\footnote{Luminosity, size and
surface brightness are related by $\mu^{e}_{\mbox{\tiny
HLR}}=M+2.5\log_{10}[2 \pi R_{\mbox{\tiny HLR}}^2]+36.57$ where
$\mu_{e}$ is the effective surface brightness, $M$ the absolute
magnitude, and $R_{\mbox{\tiny HLR}}$ the semi-major axis half-light
radius in kpc, hence the luminosity-surface brightness relation can be
readily transformed to a luminosity-size distribution and we use the
acronym LSP to indicate either.} relation for field galaxies based on
diverse datasets including: the Hubble Deep Field \citep{driver99};
the two-degree Field Galaxy Redshift Survey \citep{cross01}; the Sloan
Digital Sky Survey (\citealp{blanton01}; \citealp{shen03}); and a very
local inclination and dust corrected sample of late-type disks
\citep{dejong}. These studies consistently show that low surface
brightness is synonymous with low luminosity --- with a few notable
exceptions as typified by Malin 1 \citep{malin1} and the faint second
disk surrounding NGC5084.

To fully resolve the potential impact of surface brightness selection
effects one must consider the joint luminosity-surface brightness
distribution. This has been advocated in the past, not so much to
compensate for selection bias, but to preserve the size (or surface
brightness) information which may be of interest in its own right (see
\citealp{chol} and \citealp{sodre93} for instance). This latter point
is illustrated in Plate 1, where I show an example LF for a nearby
volume limited sample and images of the actual galaxies contributing
to the LF. Clearly much information is lost when one replaces these
images with three simple numbers. It is for these reasons --- the need
to accommodate selection bias and the desire to explore additional
parameter space --- coupled with the abundance of data that now moves
us beyond the simple LD/LF to start exploring multivariate
distributions. Here I introduce two such distributions, the
luminosity-surface brightness plane (for the reasons stated above) and
the colour-luminosity plane which is also of topical interest (e.g.,
\citealp{baldry}; \citealp{hogg} and references therein).

\begin{figure}
\includegraphics{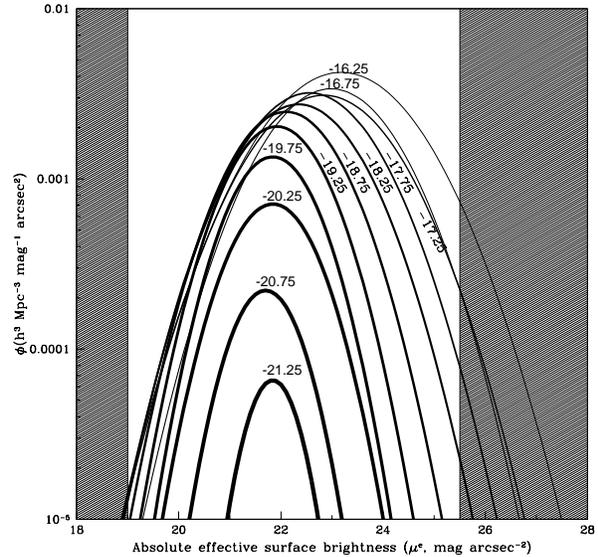}
\vspace*{75mm}
\caption{The surface brightness distribution of galaxies at various
luminosity intervals (as indicated). The curves show the Gaussian fits
to the recovered joint luminosity-surface brightness distribution of
Driver et al (2004). The shaded region denotes the limits at which
strong selection effects are likely to impact upon the observed
distributions. Generally the distribution is narrow and constant for
the brightest  galaxies (aka Freeman's Law) and then broadens towards
lower surface brightness for lower luminosity systems.
\label{fig2}}
\end{figure}

\section{Multivariate distributions}
To construct multivariate distributions requires an extensive wide
area survey ($>10^o$), to reasonable depth ($\mu_{lim} > 25 B$ mag
arcsec$^{-2}$), with reasonable resolution (FWHM $= 1''$), wavelength
coverage (e.g., some of $ub_JBVgrRiI_czJHK$\footnote{We specifically
limit ourselves to the optical/IR regime but note the existence of all
sky HI, X-ray and far-IR surveys}etc) and spectroscopic
redshifts/distances. The most notable catalogues for this purpose are:
the two-degree Field Galaxy Redshift Survey (2dFGRS; $b_jR_F$, 1800 sq
deg, 250K z's; \citealp{2dfgrs}), the Sloan Digital Sky Survey (SDSS;
$ugriz$, 10000 sq deg, 1M z's; \citealp{sdssedr}), the Millennium
Galaxy Catalogue (MGC; $B$+SDSS, 37 sq deg, 10K z's; \citealp{mgc1})
and the two Micron All Sky Survey (2MASS; $JHK$, all sky,
SDSS+2dF+6dF+MGC z's, \citealp{jarrett}). The MGC, although the
smallest in area, is also the deepest ($\mu_{lim} = 26.0 B$ mag
arcsec$^{-2}$), highest resolution (FWHM $= 1.25''$) and most complete
survey (see \citealp{mgc1}; \citealp{mgc3}; \citealp{driver04}). It
also overlaps with the other three surveys and hence provides a ``best
of all worlds'' hybrid dataset --- for example 50 per cent of the
$\sim 10,000$ MGC redshifts derive from the 2dFGRS or SDSS, extensive optical
colour coverage from SDSS, and partial near-IR coverage from
2MASS. The MGC\footnote{The MGC imaging and basic catalogues are
available from http://www.eso.org/$\sim$jliske/mgc/ (additional
catalogues including redshift, morphological and structural parameters
are available on request from spd@mso.anu.edu.au).}  contains 10,061
resolved galaxies with $12.5 <B_{MGC} < 20$ mag with 95 per cent
complete redshift coverage. All galaxies have been analysed with a
variety of software packages including SExtractor (\citealp{sex}),
GIM2D (\citealp{gim2d}) and eyeball classified to $B_{MGC} < 19$ mag.

\subsection{The luminosity-surface brightness plane}
\begin{figure}
\includegraphics{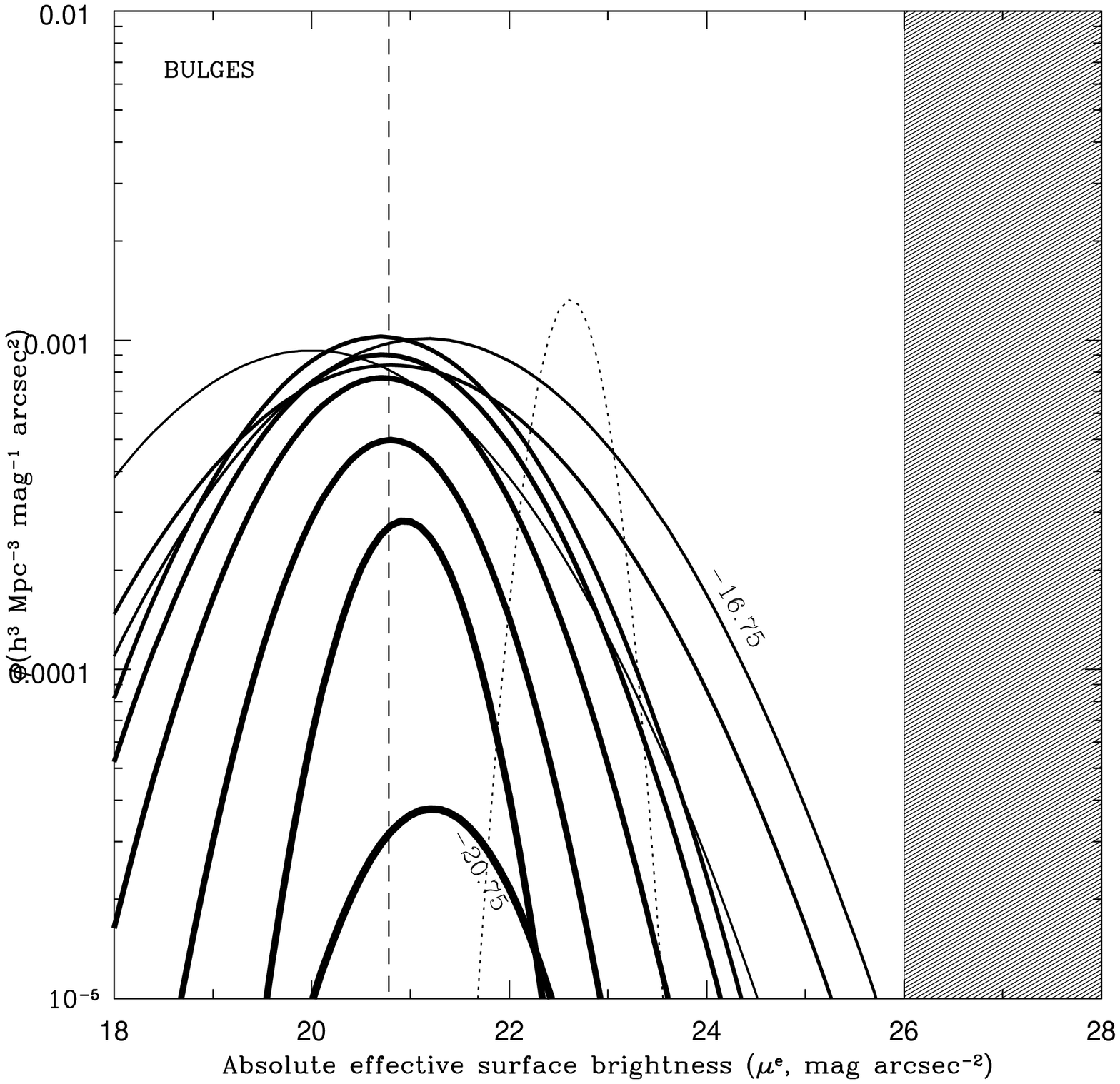}
\vspace*{75mm}
\includegraphics{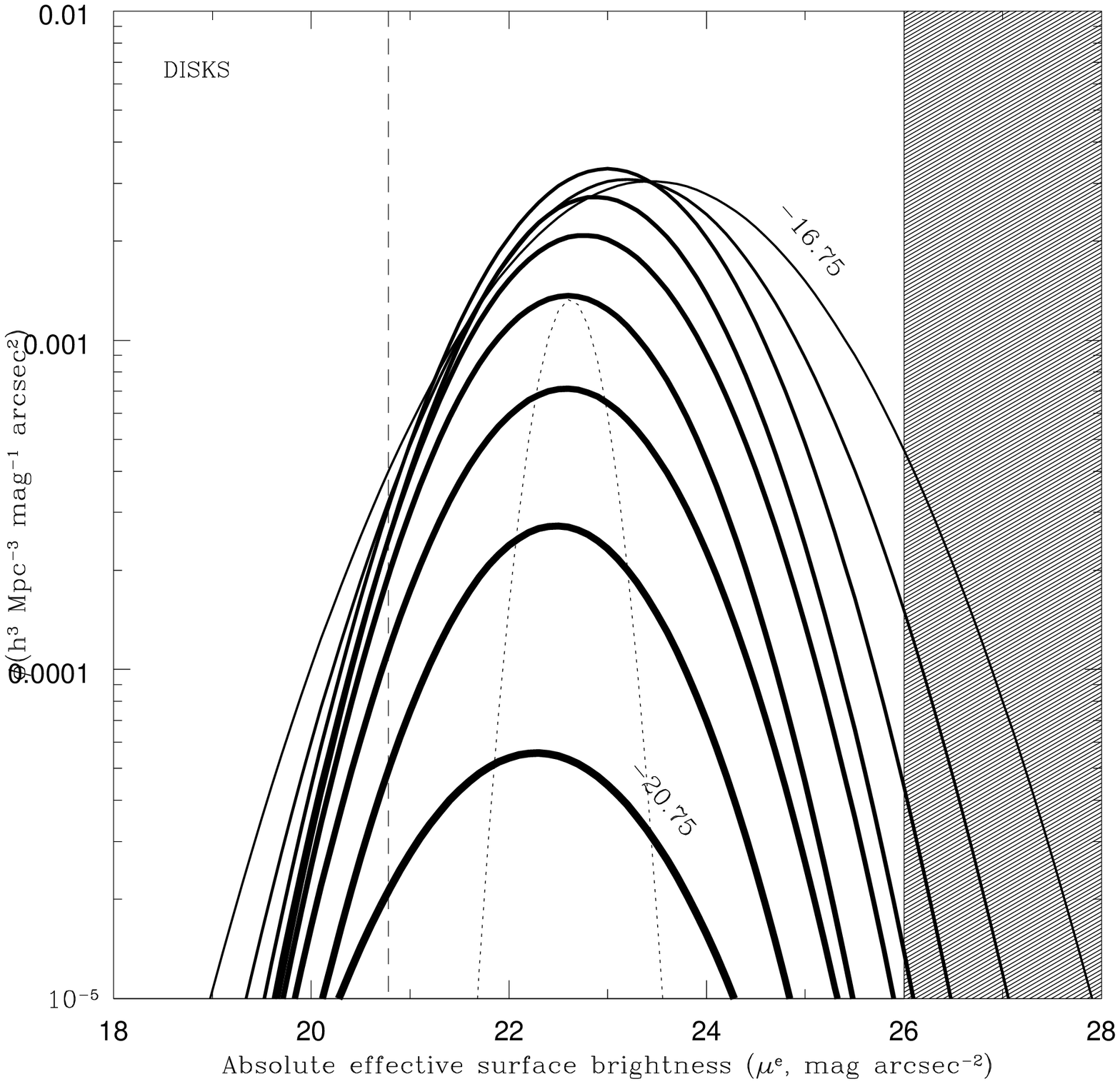}
\vspace*{75mm}
\caption{The surface brightness distributions of bulge (upper) and
disk (lower) components. The vertical dashed line shows the expected
surface brightness for spheriods at $L^*$ ($M_{B} \approx -19.6$ mag)
and the dotted curve the Freeman distribution for disks systems.  The
shaded regions show the approximate selection boundaries. (lower) as
above but for the disk components. \label{sball}}
\end{figure}

The luminosity-surface brightness plane (LSP) is of particular
interest because it enables one to compensate for both luminosity
(Malmquist bias) {\it and} surface brightness selection effects (aka
``Disney bias''). In \cite{driver04} the LSP is derived for the MGC,
which provides the most robust current estimate.  The MGC LSP analysis
used the joint luminosity-surface brightness Step-Wise Maximum
Likelihood method of \cite{sodre93} and incorporate into this tracking
of 5 selection boundaries relevant to each individual galaxy ({\it
i.e.,} maximum \& minimum observable size \& flux and minimum
observable central surface brightness for detection, see
\citealp{driver99}). An additional feature is the derivation of
individual K-corrections using the combined MGC and SDSS-DR1 colours
($uBgriz$). Fig.~\ref{fig2} shows the data as a series of Gaussian
fits across the LSP at progressive intervals of absolute magnitude.
The thicker weighted lines shows the surface brightness distribution
for the most luminous galaxies and the fainter lines for the dwarf
regime.  Two facts leap out. Firstly that the distributions are
bounded (the Gaussian fits have good $\chi^2$'s) broadening towards
lower luminosity. Secondly the peak of the distribution moves towards
lower surface brightness for lower luminosity systems. In other words
low luminosity systems apparently show greater surface brightness
diversity than giant systems. However this can also be interpreted in
terms of the Kormendy relation for spheriods (\citealp{kormendy}) and
Freeman's Law for disks (\citealp{ken}).  These two classic studies
unveiled distinct relations for the structural properties of spheroid
and disk components. The Kormendy study found that the more luminous
the spheroid the lower its central surface brightness. Conversely
Freeman's study found that all disks, regardless of luminosity, have a
constant central surface brightness of $\mu^o_{B_{\mbox{\tiny \sc
MGC}}} = 21.65 \pm 0.3$ mag arcsec$^{-2}$. The MGC result shown on
Fig.~\ref{fig2} are for the combined bulge+disk systems.  Around $L^*$
the effective surface brightness for spheriods and disks is fairly
close --- a long-time nagging coincidence. However moving towards
lower luminosity the trends for spheriods and disks diverge leading to
the broadening of the global surface brightness distribution. To
investigate further hence requires separating out these two structural
components via 2D bulge-disk decomposition. Here we use GIM2D
(\citealp{gim2d}) and Fig.~\ref{sball} shows the data of Fig.~2
subdivided by structural component.
The dotted line shows the original Freeman distribution which remains
relevant today, albeit with a far broader dispersion than originally
reported (see \citealp{ken}). 
It would
seem that galaxies consist of two principle components (presumably
formed via two mechanisms: merging and accretion/collapse ?) and to
unravel these two phases in detail must require robust bulge-disk
decompositions of extensive samples over a variety of epochs.

\subsection{The colour-luminosity plane}
\begin{figure}
\includegraphics{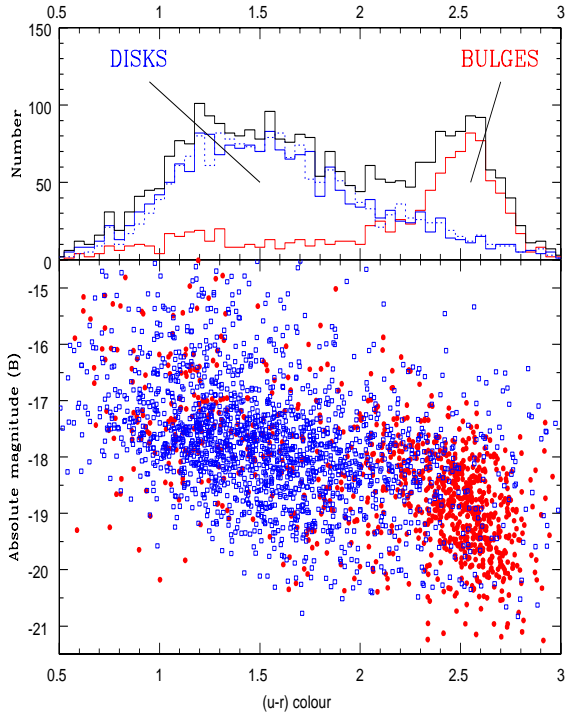}
\vspace*{105mm}
\caption{(upper) The bimodal distribution of rest-$(u-r)$ colour
(white) subdivided into bulge (red) or disk (blue) components. It is
clear that the bimodal distribution is is really a bulge-disk
dichotomy. (lower) the same data shown according to rest-($u-r$)
colour and absolute (B) magnitude. The sample is not volume corrected
but nevertheless shows that bluer systems are typically of lower
luminosity. \label{fig5}}
\end{figure}

The next most obvious key global parameter, after luminosity and
surface brightness (size) is colour, and in particular the rest
$(u-r)$ which straddles the 4000\AA -break and hence a crude indicator
of the current star-formation rate. \cite{baldry} and \cite{hogg} have
recently studied this plane extensively with SDSS data and demonstrate
clear bimodality of the colour distribution. Fig.~\ref{fig5} shows
this trend for the 10k galaxies of the MGC (using SDSS colours).
Fig.~\ref{fig5} also shows this trend for the bulge and disks
separately. To obtain the bulge colour we use the SDSS PSF magnitudes
and to obtain the disk colour and we remove the bulge colour component
from the global colour to reveal the disk colour\footnote{{\it i.e.,}
$(u-r)_{D}=-2.5\log[10^{(-0.4 u_T)}-10^{(-0.4 (u-r)_{B})}] - r_T -2.5
\log[10^{(1-B/T)}]$ where the filter subscript refers to total
Petrosian magnitude (T), disk (D) or bulge PSF magnitude (B).}. We now
see that the bimodal distribution can readily be explained in terms of
predominantly red bulges and blue disks. This component segregation
implies distinct stellar populations with distinct evolutionary
paths. Bulges must contain old stellar population and disks
intermediate or young populations. Again this follows conventional
wisdom but highlights yet further the important of bulge-disk
decomposition and the need to study the component properties of
galaxies rather than the global properties.

\section{Future prospects}
The two distributions outlined above, both suggest that the well known
bulge and disk components of galaxies follow distinct trends in both
the surface brightness (size) distribution and colour distribution.
This is of course not particularly new, however what is exciting is
our ability to quantify these distributions and trends in detail for
large statistical samples, and to extend this kind of structural
analysis to higher redshift. In particular the data resolution and
signal-to-noise of the ground-based data discussed above is comparable
to that available with the Hubble Space Telescope
(\citealp{driver95a}, \citealp{driver95b}, \citealp{driver98b},
\citealp{driver99}) and the upcoming James Webb Space Telescope. There
is nothing, other than hard diligent work, to prevent us from
quantifying the evolution of these distributions across the entire
path length of the universe. However three further issues are worth
raising:

{\bf
\noindent
1) Which wavelength is optimal for structural studies of galaxies ?

\noindent
2) How might we push back the boundaries into the dwarf regime ?

\noindent 
3) Can we connect structural measurements to the properties of the
dark matter halo ?  }

\begin{figure*}
\includegraphics{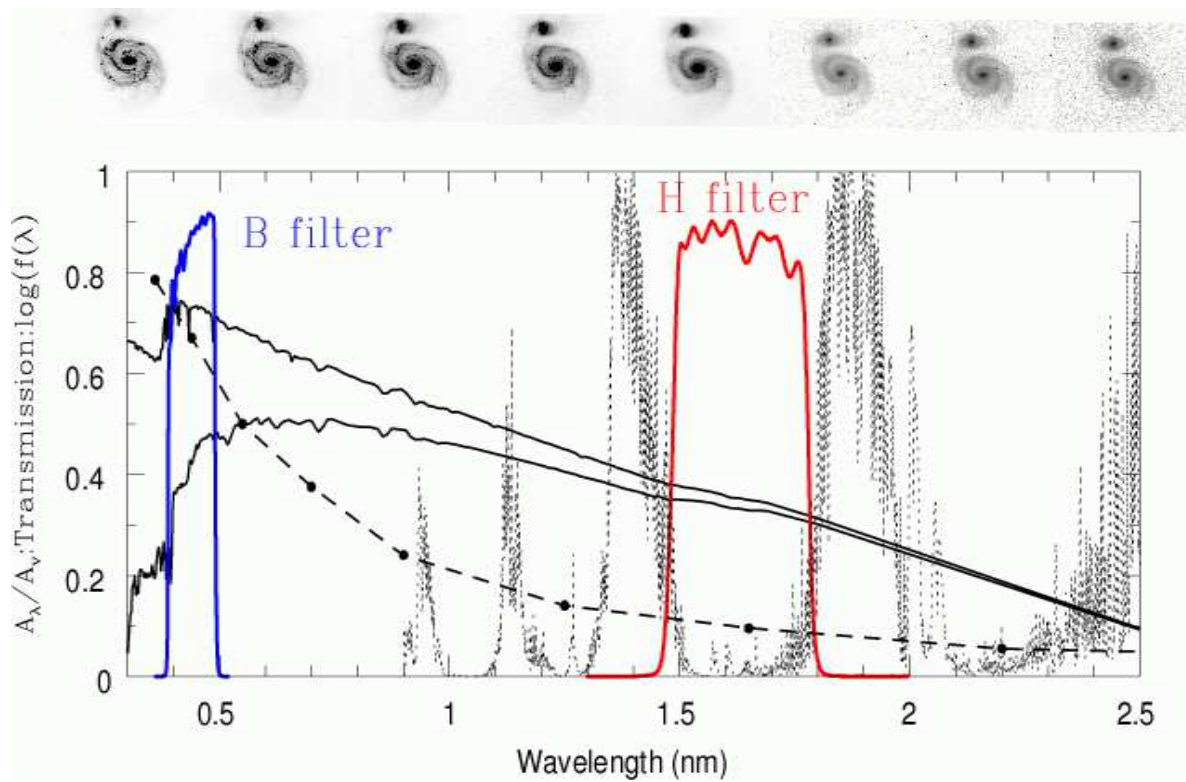}
\vspace*{100mm}
\caption{An illustration of the advantages of the near-IR. M51 images
in $UBVRIJHK$ respectively are shown along the top. The main panel
shows a spectrum of the night sky, a spectrum of a galaxy before and
after star-burst and the location of the $B$ and $H$ filters. The
extinction curve is also show. At shorter wavelengths one has to
contend with the vagaries of dust and star-formation. At longer
wavelength images are smoother, less affected by dust and
star-formation. For detailed structural analysis the longer wavelengths
are clearly optimal \label{fig6}.}
\end{figure*}

\subsection{The near-IR}
Traditionally, almost all nearby galaxy catalogues have been based on
flux-limited observations through an optical $B$ or blue bandpass
filter ($\sim 400-450$ nm), for example: the RC3; the Two-degree Field
Galaxy Redshift Survey; the Millennium Galaxy Catalogue etc. This has
been driven by technological/commercial necessity with flux detectors
typically optimised to the spectral response of the human eye
($400-800$nm). However the most physically meaningful bandpass in
which to observe a galaxy is in the near-IR (rest-$H$ band or $1.65
\mu$m). This is mostly because the stellar population that dominates
the total stellar mass --- and therefore best traces a galaxy's
gravitational potential --- is the long-lived low-mass population
which emits in the near-IR (see for example \citealp{gavazzi}). This
is most clearly demonstrated by the obviously smoother appearance of a
galaxy in the near-IR than in progressively bluer wavelengths (see
upper panel of Fig.~\ref{fig6} showing a montage of images for M51 in
a variety of filters --- the near-IR images (right most) are
significantly smoother. The flux and shape of a galaxy in the near-IR
is most dependent on the older relaxed stellar population and
therefore a better tracer of the underlying potential. Conversely the
flux and shape in the optical is linked to the young stars and
therefore dependent on the current and possibly transient
star-formation rate.  Both optical and near-IR data are important if
one wishes to understand galaxy formation and evolution. The near-IR
however appears to be the optimal filter for the investigation of the
structural properties. The other great advantage is of course the
minimisation of the impact of dust obscuration.  This is illustrated
in the main panel of Fig.~\ref{fig6} which shows the location of the
$B$ and $H$ band filters superimposed on the night sky spectrum
(dotted line), a galaxy's continuum before and after star-burst (solid
lines), and the dust attenuation curve (long dashed line). The impact
of star-formation and dust is clearly less in the near-IR. The
upcoming near-IR facilities and in particular the UKIRT/WFCAM and
VISTA will have the capabilities to provide exactly the kind of wide,
deep, high resolution data required for the comprehensive structural
analysis of nearby galaxies.

\subsection{The dwarf regime}
\begin{figure}
\includegraphics{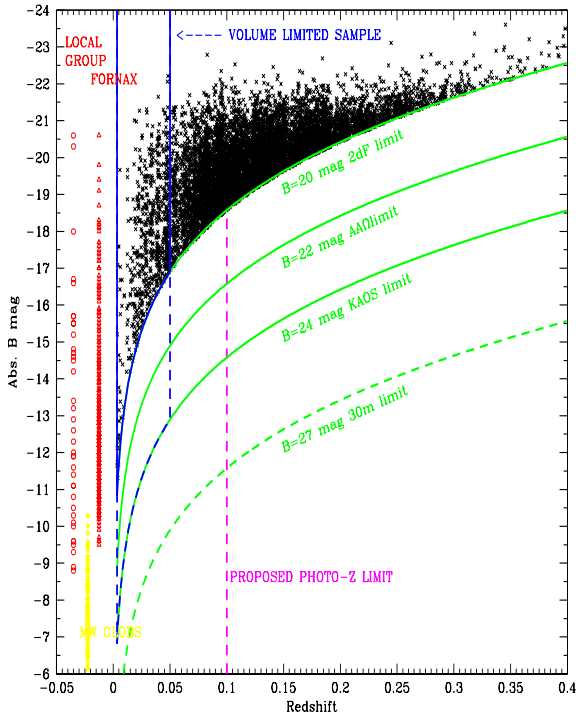}
\vspace*{105mm}
\caption{The difficulty of quantifying the faint-end of the luminosity
function is highlighted in an M v z plot such as this one. The data
points are from the MGC which is limited at $B=20$ mag. The yellow box
marks its boundary and it barely contains any volume for very low
luminosity systems. 
One possible way forward is to simply push progressively
deeper with the upcoming multiplex spectrographs (e.g., AA$\Omega$ on
the Anglo Australian Telescope or GMOS/KAOS on Gemini). The main
problem with this approach is the sheer numbers of objects. To
circumvent this one can envisage implementing a photometric redshift
cut first to pre-select candidate low-z objects.  The left side
indicates the distribution of galaxies from the local group, Fornax
and the Milky Way globular cluster distribution as
indicated. \label{fig7}.}
\end{figure}

The space density of dwarf galaxies remains
elusive. Figs.~\ref{fig1}~\&~\ref{fig2} shows that the MGC can only
sample with credibility to $M_{B} \approx -16$ mag at which point both
limiting statistics and the high and low surface brightness selection
limits bite (see \citealp{driver04}). Fig.~\ref{fig7} illustrates this
by showing the MGC galaxies on an absolute magnitude versus redshift
plot. The data are of course bounded by the $B=20$ mag limit which
highlights the rapidly diminishing volume observed for low luminosity
systems. One way to overcome this is to simply conduct ever deeper
redshift surveys (as indicated on the figure). However this has a
diminishing return as the number of galaxies one must observe to find
one low luminosity system becomes unreasonable. One possible way
forward is to use photometric redshifts to pre-select low-z candidates
and then follow-up only these systems. However the accuracy of
photometric redshifts at low z is poor (although improved by near-IR
colours, see \citealp{bolzonella} for example). To overcome the
surface brightness selection limits (both high and low) the source
data must be improved to probe to very high resolution ($FWHM <
0.5''$) and very deep isophotes ($\mu_B >> 26$ mag arcsec$^{-2}$) over
wide areas ($30+$ sq deg). No such survey exists but facilities such
as SUBARU/SUPrime and Magellan/IMACS just about have the capability to
achieve such a survey.  The alternative method is to observe the very
local galaxy population (i.e., the Local Sphere of Influence, defined
as that within 10 Mpc) and obtain direct distance measurements rather
than redshifts.

\subsection{Physics of the LSP ?}
\begin{figure}
\includegraphics{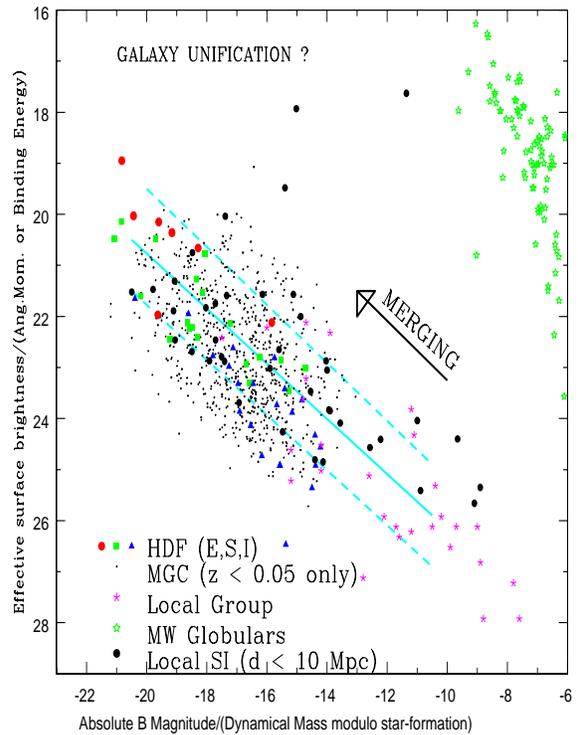}
\vspace*{105mm}
\caption{A summary of available LSP data drawn from a variety of
sources. The red line marks the credibly mapped area and the cyan line
shows the expectation from de Jong \& Lacey (2000). This appears to
follow the data remarkably well. \label{fig8}}
\end{figure}

The LSP may have the potential to connect key observables (luminosity
and size) to the fundamental underlying physical properties of bulge
and disk systems (mass and angular momentum). In various studies of
the formation of disk systems, ({\it e.g.,} \citealp{fall},
\citealp{dss97}, \citealp{mo}) the dimensionless spin parameter
($\lambda = J|E^{\frac{1}{2}}|G^{-1}M_{\mbox{\tiny
halo}}^{-\frac{5}{2}}$, \citealp{peebles}) is directly related to the
scale-length of the disk. The spin parameter reflects how close the
halo is to a rotationally supported system and is a key parameter
monitored by the numerical simulations (see \cite{steinmetz};
\citealp{cole}; \citealp{vitvitska}; Maller, Dekel \& Sommerville 2002
for example). The pivotal idea (here echoing the toy model of
\citealp{dejong}) starts with the premise that the baryons are coupled
to the dark matter halo, because of this the luminosity (generated by
the baryons in the form of stars) can be related to the systemic mass
and the rotation of the stars/gas can be related to the systemic
angular momentum. Given this premise, which is intimated by the
Tully-Fisher relation, one can analytically relate $\lambda$ to
luminosity and surface brightness (or size): $\lambda \propto
\Sigma_{eff}^{-\frac{1}{2}} L^{-\frac{\gamma}{3}+\frac{1}{2}}$ (from
\citealp{dejong}), where $\Sigma_{eff}$ is the effective surface
brightness, $L$ is the intrinsic luminosity in some filter and
$\gamma$ is the dependence of luminosity on the mass-to-light ratio
(equal to 0.69 in $B$ or 1.00 in $H$ \citealp{gavazzi}). Numerical
simulations consistently find that the distribution of the spin
parameter is a log Normal distribution which is globally preserved
through hierarchical merging (see for example \citealp{vitvitska})
this yields: $\Sigma_{eff}=L_{B}^{0.54}$ or $\mu_{eff}=0.54
M_{B}$. Hence the gradient of any luminosity-surface brightness
relation bears upon the relation between mass and light and the
dispersion upon the breadth of the spin distribution. Fig.~\ref{fig8}
shows the $B$-band LSP for a variety of samples as indicated (LG,
\citealp{mateo}; HDF, \citealp{driver99}; MGC, \citealp{driver04}; MW
GCs, Harris et al (priv. comm); Local Sphere of Influence,
\citealp{jerjen}; LSBGs, \citealp{deblok}). The solid lines show the
approximate expectation as argued above and show remarkable agreement
with the data - in detail the observed size distribution is marginally
narrower than simulations predict (see \cite{driver04}). It is also
worth noting that systems which form via merging (i.e., bulges) and
via accretion (i.e., disks) are also predicted to show distinct
$\lambda$ distributions (see for example \citealp{vitvitska};
\citealp{maller}). At the moment far more data and detailed
simulations are required however this connection is clearly promising
and could ultimately result in a galaxy equivalent to the
Hertzsprung-Russell diagram, allowing a meeting ground between
numerical simulations and survey observations.

\section{Summary}
The galaxy luminosity function is not the only fruit, and with the
many legacy datasets becoming available the time is now ripe to move
beyond the LF and explore multivariate distributions. Here I've
presented two: the Luminosity surface brightness (size) plane (LSP);
and the colour luminosity plane (CLP). Both planes show that disks and
bulges form distinct but overlapping distributions presumably
indicating secular evolution of these components, i.e., two mechanisms
and two timescales. This finding argues for the community to move away
from global measurements and start to measure the properties of these
distinct components independently. We argue that this is best done in
the near-IR and should be a key focus of upcoming IR facilities such as
VISTA (low-z) \& JWST (high-z). Perhaps most important of all the LSP
appears to provide a direct meeting ground to the numerical
simulations. This last point is by far the most important as its from
the cross-talk between simulations and observations that real insight
into the processes of galaxy evolution and formation will come.

\section{Acknowledgments}
I am indebted to my colleagues on the MGC and to the hard work of
those that have slaved on the superb 2dFGRS, SDSS and 2MASS facilities
and databases. These data are changing the way we view galaxies in a
profound way.  Finally I'd like to thank the organisers for a very
enjoyable meeting and support from PPARC as a visiting fellow to the
University of Bristol where this article was revised.


\onecolumn

\begin{figure}
\includegraphics{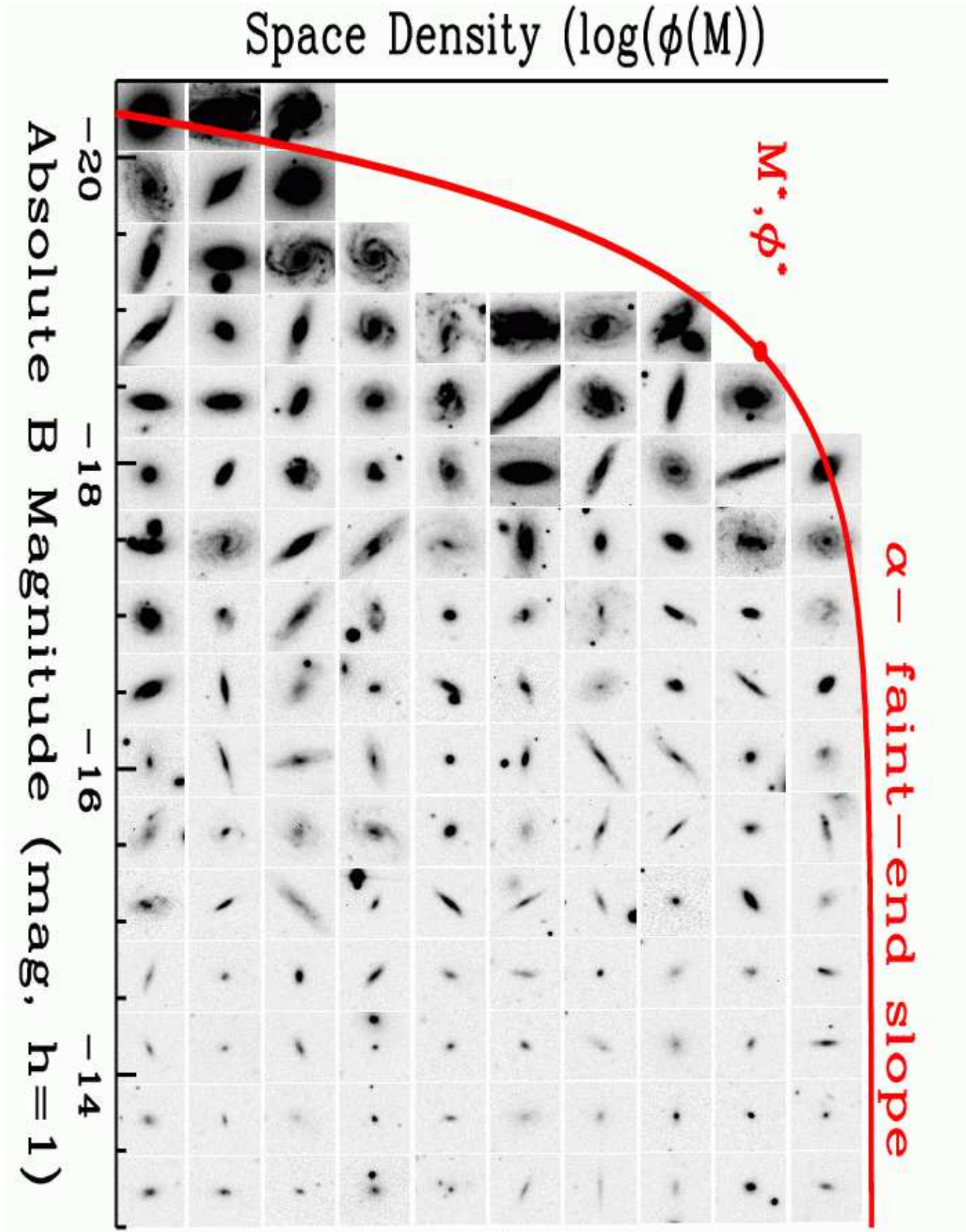}
\vspace*{227mm}
\caption{PLATE1: The global Galaxy Luminosity Function (red line) condenses
the available information of galaxies (images) into three crucial
numbers: the characteristc luminosity ($M^*$); the absolute
normalisation ($\phi^*$); and the faint-end slope ($\alpha$). Although
the schechter parameterisation is more often that not a remarkably
good fit, one cannot help but feel that too much important infomation
may have been lost, for instance the sizes and bulge-to-total
parameters. \label{plate1}}
\end{figure}

 \label{lastpage}


\begin{thebibliography}{}

\bibitem[\protect\citeauthoryear{Andreon \&
Culliandre}{2002}]{andreon} Andreon S., Culliandre, J,-C., 2002, ApJ,
569, 144

\bibitem[\protect\citeauthoryear{Baldry et al.}{2003}]{baldry} Baldry
Glazebrook K., Brinkmann J., Ivezic Z., Lupton R.H.,
Nichol R.C., Szalay A.S., 2004, ApJ, 600, 681

\bibitem[\protect\citeauthoryear{Beijersbergen et
al}{2002}]{beijer2002} Beijersbergen, M., Hoekstra, H., van Dokkum,
P.G., \& van der Hulst, T., 2002, MNRAS, 329, 385

\bibitem[\protect\citeauthoryear{Bertin \& Arnouts}{1996}]{sex} Bertin
E., Arnouts S., 1996, A\&AS, 117, 393

\bibitem[\protect\citeauthoryear{Blanton et al.}{2001}]{blanton01}
Blanton M.R., et al. 2001, AJ, 121, 2358

\bibitem[\protect\citeauthoryear{Blanton et al.}{2003a}]{blanton03}
Blanton M.R., et al. 2003, ApJ, 592, 819

\bibitem[\protect\citeauthoryear{Bolzonella, Miralles \& Pell\'o}{2000}]
{bolzonella}
Bolzonella M., Mirralles, J.-M., Pell\'o R., 2000, A\&A, 363, 476

\bibitem[\protect\citeauthoryear{Bothun et al}{1988}]{malin1} Bothun G.D.,
Impey C., Mould J., Malin D., 1987, MNRAS, 94, 23

\bibitem[\protect\citeauthoryear{Cho\l oniewski}{1985}]{chol} Cho\l
oniewski J., 1985, MNRAS, 214, 197

\bibitem[\protect\citeauthoryear{Cole \& Lacey}{1996}]{cole} Cole S.,
Lacey C., 1996, MNRAS, 281, 716

\bibitem[\protect\citeauthoryear{Colless et al.}{2001}]{2dfgrs}
Colless M., et al., 2001, MNRAS, 328, 1039

\bibitem[\protect\citeauthoryear{Cross et al.}{2001}]{cross01} Cross
N.J.G., et al., 2001, MNRAS, 324, 825

\bibitem[\protect\citeauthoryear{Cross \& Driver}{2002}]{cross02}
Cross N.J.G., Driver S.P., 2002, MNRAS, 329, 579

\bibitem[\protect\citeauthoryear{Cross et al.}{2004}]{mgc3} Cross
N.J.G., et al.  2004, MNRAS, in press

\bibitem[\protect\citeauthoryear{Croton et al.}{2004}]{croton} Croton, D.,
et al.  2004, MNRAS, submitted (astro-ph/0407537)

\bibitem[\protect\citeauthoryear{Dalcanton}{1998}]{dalcanton}
Dalcanton J.J., 1998, ApJ, 495, 251

\bibitem[\protect\citeauthoryear{Dalcanton et. al.}{1997}]{dss97}
Dalcanton J.J., Spergel, D., Summers, F.J., 1997, ApJ, 482, 659

\bibitem[\protect\citeauthoryear{de Blok, van der Hulst \&
Bothun}{1995}]{deblok} de Blok, E., van der Hulst T., Bothun G.D.,
1995, MNRAS, 274, 235

\bibitem[\protect\citeauthoryear{de Jong \& Lacey}{2000}]{dejong} de
Jong R.S., Lacey C., 2000, ApJ, 545, 781

\bibitem[\protect\citeauthoryear{De Propris et al}{2003}]{rdp2003}
De Propris, R.,  et al 2003, MNRAS, 342, 725

\bibitem[\protect\citeauthoryear{Disney}{1976}]{disney} Disney M.J.,
1976, Nature, 263, 573

\bibitem[\protect\citeauthoryear{Driver et al}{2004}]{driver04}
Driver S.P., Liske J., Cross N.J.G., De Propris R., Allen P.D., 2004,
MNRAS, submitted

\bibitem[\protect\citeauthoryear{Driver \& De
Propris}{2003}]{driver03} Driver S.P., De Propris R., 2003, Ap\&SS,
285, 175

\bibitem[\protect\citeauthoryear{Driver et al}{2003}]{a868} Driver
S.P., Odewahn S.C., Echevarria L., Cohen S.H., Windhorst R.A.,
Phillipps S., Couch W.J., 2003, AJ, 126, 2662

\bibitem[\protect\citeauthoryear{Driver et al}{1994}]{a963}
Driver S.P., Phillipps S., Morgan I., Davies J.I., Disney M.J., 1994,
MNRAS, 268, 393

\bibitem[\protect\citeauthoryear{Driver et al}{1998a}]{driver98}
Driver S.P., Couch W.J., Phillipps S., 1998, MNRAS, 301, 369

\bibitem[\protect\citeauthoryear{Driver}{1999}]{driver99} Driver S.P.,
1999, ApJ, 526, 69

\bibitem[\protect\citeauthoryear{Driver et al}{1995a}]{driver95a}
Driver S.P., Windhorst R.A., Griffiths R.E., 1995, ApJ, 453, 48

\bibitem[\protect\citeauthoryear{Driver et al}{1995b}]{driver95b}
Driver S.P., Windhorst R.A., Ostrander E.J., Keel W.C, Griffiths R.E.,
Ratnatunga K.U, ApJL, 449, 23

\bibitem[\protect\citeauthoryear{Driver et al}{1998b}]{driver98b}
Driver S.P., Fernandez-Soto A., Couch W.J., Odewahn S.C., Windhorst R.A.,
Phillipps S., Lanzetta K., Yahil A., 1998, 496, 93

\bibitem[\protect\citeauthoryear{Fall \& Efstathiou}{1980}]{fall} Fall
M., Efstathiou G., 1980, MNRAS, 193, 189

\bibitem[\protect\citeauthoryear{Freeman}{1970}]{ken} Freeman K.C.,
1970, AJ, 160, 811

\bibitem[\protect\citeauthoryear{Gavazzi, Pierrini \& Boselli
}{1996}]{gavazzi} Gavazzi G., Pierrini D., Boselli A., 1996, 312, 397

\bibitem[\protect\citeauthoryear{Hogg et al.}{2004}]{hogg} Hogg D.W.,
et al, 2004, ApJL, 601, 29

\bibitem[\protect\citeauthoryear{Impey \& Bothun}{1997}]{impey97}
Impey C., \& Bothun G.D., 1997, ARA\&A, 35, 267

\bibitem[\protect\citeauthoryear{Impey \& Bothun}{1988}]{bothun} Impey
C., Bothun G.D., 1988, ApJ, 330, 634

\bibitem[\protect\citeauthoryear{Jarrett et al}{2003}]{jarrett} Jarrett,
Jarrett T.H., Chester T., Cutri R., Schneider S.E., Huchra J.P. 2003, AJ,
125, 525

\bibitem[\protect\citeauthoryear{Jerjen, Binggeli \& Freeman}{2000}]{jerjen}
Jerjen, H. Binggeli, B., Freeman, K.C., 2000, ApJ, 119, 593

\bibitem[\protect\citeauthoryear{Kormendy}{1977}]{kormendy} Kormendy
J., 1977, ApJ, 218, 333

\bibitem[\protect\citeauthoryear{Liske et al.}{2003}]{mgc1} Liske J.,
Lemon D., Driver S.P., Cross N.J.G., Couch W.J., 2003, MNRAS, 344, 307

\bibitem[\protect\citeauthoryear{Norberg et al.}{2002}]{norberg02}
Norberg P., et al. 2002, MNRAS, 336, 907

\bibitem[\protect\citeauthoryear{Maller et al.,}{2002}]{maller}
Maller A.H., Dekel A.,  Sommerville R., 329, 423

\bibitem[\protect\citeauthoryear{Mateo}{1998}]{mateo} Mateo M., 1998,
ARA\&, 36, 435

\bibitem[\protect\citeauthoryear{Mo et al.}{1998}]{mo} Mo S., Mao
H.J., White S.D.M., 1998, MNRAS, 297, 71

\bibitem[\protect\citeauthoryear{Mobasher et al}{2004}]{mobasher2003}
Mobasher, B., et al., 2004, MNRAS, in press (astro-ph/0301047)

\bibitem[\protect\citeauthoryear{Peebles}{1969}]{peebles} Peebles J.,
1969, ApJ, 155, 393

\bibitem[\protect\citeauthoryear{Press \& Schechter}{1974}]{PS74}
Press, W., Schechter, P.J., 1974, ApJ, 187, 425

\bibitem[\protect\citeauthoryear{Phillipps et al}{1998}]{phillipps}
Phillipps S., Driver S.P., Couch W.J., Smith R.M., 1998, ApJ, 498, 119

\bibitem[\protect\citeauthoryear{Pracey et al}{2004}]{pracey} Pracey, M.,
et al , MNRAS, in press

\bibitem[\protect\citeauthoryear{Schechter}{1976}]{schechter} Schechter P.,
1976, ApJ, 203, 297

\bibitem[\protect\citeauthoryear{Shen et al.}{2003}]{shen03} Shen S.,
Mo H.J., White S.D.M., Blanton M.R., Kauffmann G., Voges W., Brinkmann
J., Csabai I., 2003, MNRAS, 343, 978

\bibitem[\protect\citeauthoryear{Simard et al.}{2002}]{gim2d} Simard
L., et al., 2002, ApJS, 142, 1

\bibitem[\protect\citeauthoryear{Smith et al}{1997}]{smith} Smith R.M.,
 Driver S.P., Phillipps S., Couch W.J., 19997, MNRAS, 287, 415

\bibitem[\protect\citeauthoryear{Sodr\'e \& Lahav}{1993}]{sodre93}
Sodr\'e L., Jr., Lahav O., 1993, MNRAS, 260, 285

\bibitem[\protect\citeauthoryear{Sprayberry et al.}{1997}]{spray}
Sprayberry D., Impey C., Irwin M.J., Bothun G.D., 1997, ApJ, 482, 104

\bibitem[\protect\citeauthoryear{Steinmetz \& Bartelmann}{1995}]{steinmetz}
Steinmetz M., Bartelmann M., 1995, MNRAS, 272, 570

\bibitem[\protect\citeauthoryear{Stoughton et al.}{2002}]{sdssedr}
Stoughton C., et al. 2002, AJ, 123, 485

\bibitem[\protect\citeauthoryear{Trentham}{1998}]{trentham} Trentham
N., 1998, MNRAS, 293, 71

\bibitem[\protect\citeauthoryear{Trentham \&
Hodgkin}{2002}]{trent2002} Trentham, N., Hodgkin, S., 2002, MNRAS,
333, 423

\bibitem[\protect\citeauthoryear{Vitvitska et al.}{2002}]{vitvitska}
Vitvitska M., Klypin A.A., Kravtsov A.V., Wechsler R.H., Primack
J.R., Bullock J.S., 2002, ApJ, 581, 799

\end{thebibliography}
\end{document}